\documentclass[12pt,preprint]{aastex}

\def\hi{\ifmmode {\rm H}\,{\sc i}~ \else H\,{\sc i}~\fi}

\def\chandra {{\it Chandra}}
\def\xmm {{\it XMM}}
\def\xmmnewton {{\it XMM--Newton}}
\def\fuse {{\it FUSE}}
\def\cvi {\ion{C}{6}}
\def\nvi {\ion{N}{6}}

\def\nvii {\ion{N}{7}}
\def\nvi {\ion{N}{6}}

\def\ovii {\ion{O}{7}}

\shorttitle{XMM Observations of Mrk 421}
\shortauthors{Williams et al.}

\begin{document}


\title{\boldmath{\it XMM--Newton} View of the $z>0$ Warm--Hot
Intergalactic Medium Toward Markarian 421}


\author{Rik J. Williams\altaffilmark{1}, 
        Smita Mathur\altaffilmark{1},
	Fabrizio Nicastro\altaffilmark{2,3,4},
	Martin Elvis\altaffilmark{2}}
\altaffiltext{1}{Department of Astronomy, The Ohio State University, 
                 140 West 18th Avenue, Columbus OH 43210, USA}
\altaffiltext{2}{Harvard--Smithsonian Center for Astrophysics, Cambridge,
                 MA, 01238, USA}
\altaffiltext{3}{Instituto de Astronom\'ia, Universidad Aut\'onomica de
                 M\'exico, Apartado Postal 70-264, Ciudad Universitaria,
                 M\'exico, D.F., CP 04510, M\'exico}
\altaffiltext{4}{Osservatorio Astronomico di Roma, Istituto Nazionale di
                 Astrofisica, Italy}
\email{williams,smita@astronomy.ohio-state.edu}


\begin{abstract}
The recent detection with \chandra\ of two warm--hot intergalactic
medium (WHIM) filaments toward Mrk 421 by Nicastro et al.~provides
a measurement of the bulk of the ``missing baryons'' in the 
nearby universe.  Since Mrk 421 is a bright X-ray source, it is also frequently
observed by the \xmmnewton\ Reflection Grating Spectrometer (RGS) 
for calibration purposes.  Using all available archived
\xmm\ observations of this source with small pointing offsets ($<15$\arcsec), we
construct the highest--quality \xmm\ grating spectrum of Mrk 421 to date
with a net exposure time (excluding periods of high background flux)
of 437~ks and $\sim 15000$ counts per resolution element at 21.6\,\AA,
more than twice that of the \chandra\ spectrum.  Despite the long exposure 
time neither of the two intervening
absorption systems is seen, though the upper limits derived are 
consistent with the \chandra\ equivalent width measurements.  This
appears to result from (1) the larger number of narrow instrumental features
caused by bad detector columns, (2)
the degraded resolution of XMM/RGS as compared to the \chandra/LETG, 
and (3) fixed pattern noise at $\lambda \ga 29$\,\AA. 
The non--detection of the WHIM absorbers by XMM is thus fully consistent
with the \chandra\ measurement.

\end{abstract}




\section{Introduction}
Most of the baryons that comprise 4\% of the mass--energy budget
of the universe are found in the intergalactic medium (IGM), primarily 
appearing as the Lyman--alpha ``forest'' in high--redshift
quasar spectra \citep{weinberg97}.  At more recent times ($z\la 2$) 
the process of structure formation has
shock--heated the IGM to temperatures of $\sim 10^{5-7}$\,K, thus
rendering the hydrogen nearly fully ionized and producing (at most)
broad, extremely weak Lyman--alpha absorption 
\citep[e.g.][]{sembach04,richter04}.  Known as the warm--hot intergalactic 
medium (WHIM), this phase is predicted to contain roughly half of the 
baryonic matter at low redshifts \citep{cen99,dave01}. Its extremely low 
density (typically $\delta\sim 10-100$) 
precludes the detection of WHIM thermal or line emission with current
facilities, so the only way to directly measure these ``missing'' baryons
is through far--ultraviolet and X-ray spectroscopic measurements of
absorption lines from highly ionized heavy elements 
\citep{perna98,hellsten98,fangb02}.  

Several early attempts to detect these intervening WHIM absorbers in X-rays
\citep[e.g.][]{fang02,mathur03,mckernan03} and more recent surveys
\citep{fang05} yielded only tentative detections
at best.  Intervening Lyman--alpha \citep{shull96,sembach04} and \ion{O}{6} 
\citep[e.g.][]{savage02} absorbers had also been seen in \fuse\ and HST 
quasar spectra, but their ionization states and possible galactic halo origins
\citep[e.g.][]{tumlinson05} are quite uncertain.  These uncertainties
are largely mitigated with the recent detection by 
\citet{nicastro05a,nicastro05b} 
of two X-ray absorption systems at $z=0.011$ and $z=0.027$ along the 
line of sight to the blazar Mrk 421.  These filaments account for
a critical density fraction of 
$\Omega_{\rm WHIM}=0.032^{+0.042}_{-0.021}$, fully consistent with the mass 
of the missing baryons in the local universe (albeit with large 
uncertainties).  While future proposed missions such as \emph{Constellation--X},
\emph{XEUS}, or \emph{Pharos} (Nicastro et al., in preparation) will be able to 
measure $\Omega_{\rm WHIM}$ to far greater precision with
detections of numerous weaker X-ray forest lines, the 
\citet[][hereafter N05]{nicastro05a} results present a key 
early confirmation of numerical predictions using observational
capabilities that are \emph{currently} within our grasp --- hence any
test of their correctness is of great importance.

Although each of these two absorption systems was detected with high confidence
through multiple redshifted X-ray absorption lines, the \emph{individual}
absorption lines were generally quite weakly detected, mostly at the
$2-4\sigma$ level.  Moreover, while they employed high--quality \chandra\
and \fuse\ data taken during exceptionally bright outbursts of Mrk 421, 
the many archived \xmmnewton\ observations of this source 
were not included in the analysis.  With roughly twice the effective
area of \chandra/LETG, \xmm/RGS is in principle superior for X-ray 
grating spectroscopy 
between $\sim 10-40$\,\AA; however, its slightly worse resolution 
(approximately 60\,m\AA\ FWHM, versus 50\,m\AA\ for \chandra/LETG), higher
susceptibility to background flares, and multitude of narrow instrumental 
features can present serious complications for WHIM searches.  

Independent confirmation of the \chandra\ results with a separate 
instrument like \xmm\ is thus important.  While some groups have searched
for WHIM features in a limited number of \xmm\ Mrk 421 spectra
\citep[e.g.,][]{ravasio05}, a complete and systematic analysis has yet
to be performed.  Here we present a search for $z>0$ WHIM features employing
all ``good'' observations of Mrk 421 available in the \xmm\ 
archive, and a comparison of these results to those presented by N05.

\section{Data Reduction and Measurements}
We searched the \xmm\ archive for all Mrk 421 Reflection Grating
Spectrometer (RGS) data.  Although
31 separate observations were available, 16 had pointing offsets
$\Delta\theta \ga 60$\arcsec\ while the rest were offset by less than 
15\arcsec.  Since spectral resolution and calibration quality can degrade at
large offsets, we only included
those with $\Delta\theta < 15$\arcsec.  One extremely short observation
(0158971101, with $t_{\rm exp}=237$\,s) was also excluded to simplify the
data reduction process.  Using the standard \xmm\ Science Analysis System
version 6.5.0 
routines\footnote{See \url{http://xmm.vilspa.esa.es/sas}}, RGS1 
light curves were built for the remaining 
14 ``good'' observations (see Table~\ref{tab_log}), and the spectra were 
reprocessed to exclude periods of high background levels and coadded.  
These combined, filtered  
RGS1 and RGS2 spectra have effective exposure times of $\sim 440$\,ks and 
over $9\times 10^6$
combined RGS1$+$RGS2 first--order counts between $10-36$\,\AA\ with 
$\sim 15000$ counts per 0.06\,\AA\ resolution element in RGS1 near 21\,\AA, 
over twice that in the N05 Mrk~421 \chandra\ spectrum.

We first used the spectral fitting program 
\emph{Sherpa}\footnote{\url{http://cxc.harvard.edu/sherpa/}} to fit
a simple power law plus Galactic foreground absorption model to the
RGS1 and RGS2 data; however, at such high spectral quality the RGS
response model does not appear to be well--determined, and large residuals 
remained.  For line measurements, we thus only considered
$\sim 2$\,\AA\ windows around each wavelength of interest, using a power
law to independently model
the RGS1 and RGS2 continua within each window
and excluding the strongest narrow detector features (with typical
widths of 70\,m\AA\ or less).  None of the intervening absorption
lines were apparent through a visual inspection of the \xmm\ spectrum, though
several of the $z=0$ lines reported by \citet{williams05} could be seen
clearly.  

A narrow Gaussian absorption line (FWHM$=5$\,m\AA)
was included in the model for each 
line measurement or upper limit reported by N05.  When convolved with
the RGS instrumental response these absorption lines appeared broadened
to the RGS line spread function (LSF) widths \citep[typically 
FWHM$=60-70$\,m\AA;][]{denherder01}.  The $2\sigma$  
upper limits on all equivalent widths were then calculated (allowing the
central line wavelengths to vary within the $1\sigma$ errors reported
by N05).  Since the shapes of the RGS1 and RGS2 instrumental responses
are quite different, these limits were calculated using both a joint
fit to the RGS1$+$RGS2 spectra as well as the individual RGS1 and RGS2
spectra.  It should be noted that wherever one RGS unit is unusable,
the total response is effectively halved, at which point it has a similar
effective area to \chandra/LETG.   The resulting equivalent width limits 
are listed in Table~\ref{tab_ew}.

\section{Discussion}
\begin{figure}  
\plotone{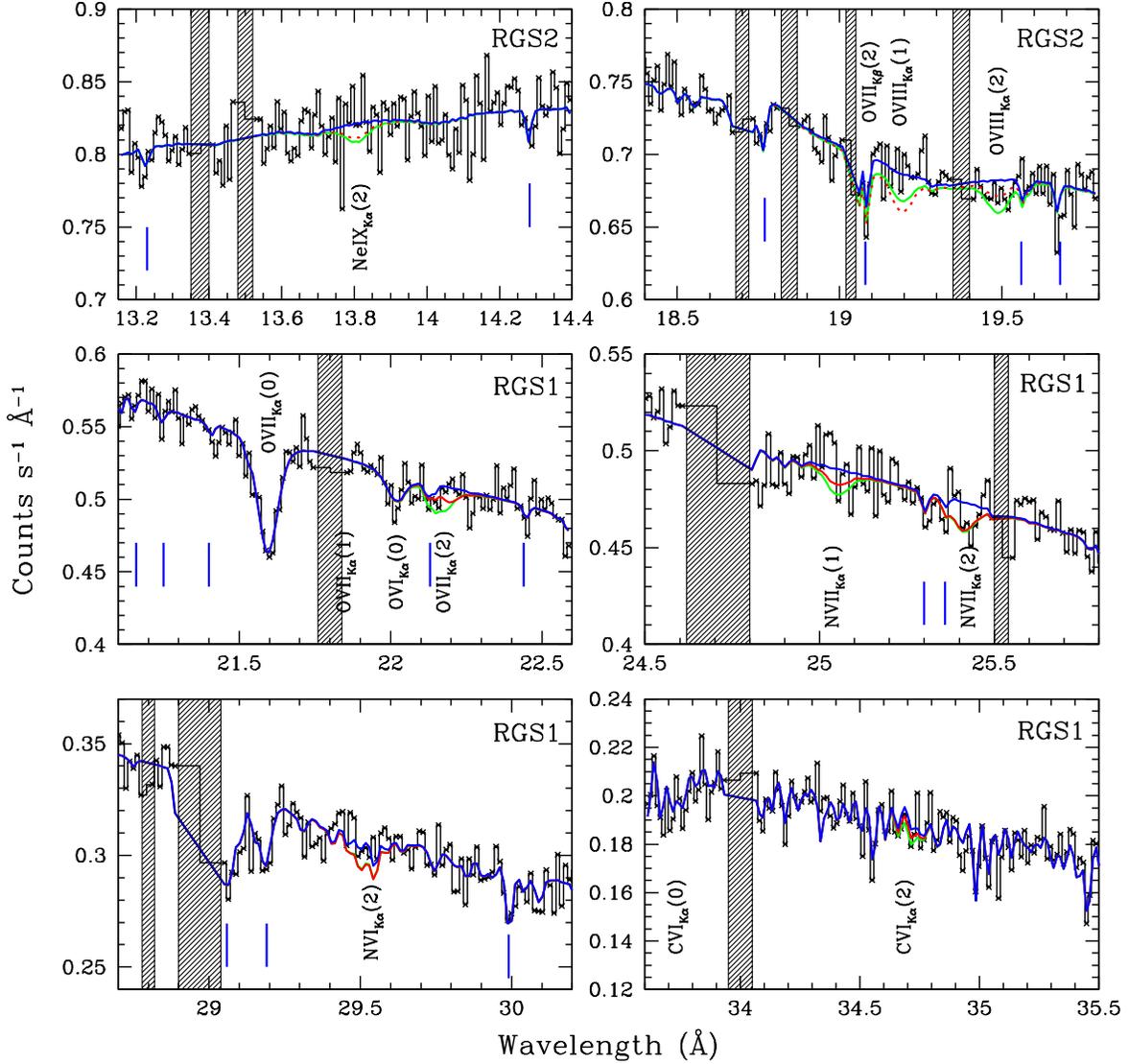}
\caption{Portions of the XMM--Newton RGS spectrum of Mrk 421 (black histogram).
The continuum fit is shown by the solid blue
line, the green line depicts the $2\sigma$ upper limit on each measured
species from XMM, and the red lines show the results of N05a
(where their $3\sigma$ upper limits are shown as dashed lines, while
solid lines indicate best--fit measurements).  All N05a absorption line 
positions are labeled, with (1) and (2) denoting $z=0.011$, and
$z=0.027$ lines respectively.  Significant $z=0$ lines 
are included in the continuum fit for consistency and labeled with (0), but not
discussed further here.  Regions that were excluded from the fit
due to chip gaps and detector features are shaded; weaker instrumental
features are marked with vertical blue ticks.  Although joint
fits using RGS1 and RGS2 were performed whenever possible, for display
purposes only one or the other (indicated in the upper right--hand corner
of each plot) is shown for each spectral region.
\label{fig_xmmspec}}
\end{figure}

Figure~\ref{fig_xmmspec} shows the spectral windows used to determine
upper limits on the N05 measured lines, with the data (black), continuum fit
(blue), \chandra\ measurements and limits (N05; red solid and dotted lines
respectively), and \xmm\ limits (green) overplotted.
In all cases, the N05 measurements (or $3\sigma$ upper
limits) appear to be consistent with the 
$2\sigma$ upper limits we have derived directly from the \xmm\ data,
as shown in the figure and listed in Table~\ref{tab_ew}. 
The \ovii\ line at $z=0.027$
looks as though it might be visible in the spectrum, but this is most
likely due to the weak instrumental feature at $\sim 22.1$\,\AA.
For two lines (\nvii\ and \nvi\ at
$z=0.027$) the \xmm\ $2\sigma$ upper limits are approximately equal to
the N05 best--fit measurements, but since the N05 values are quite uncertain
this result is still consistent.  

\begin{figure}
\plotone{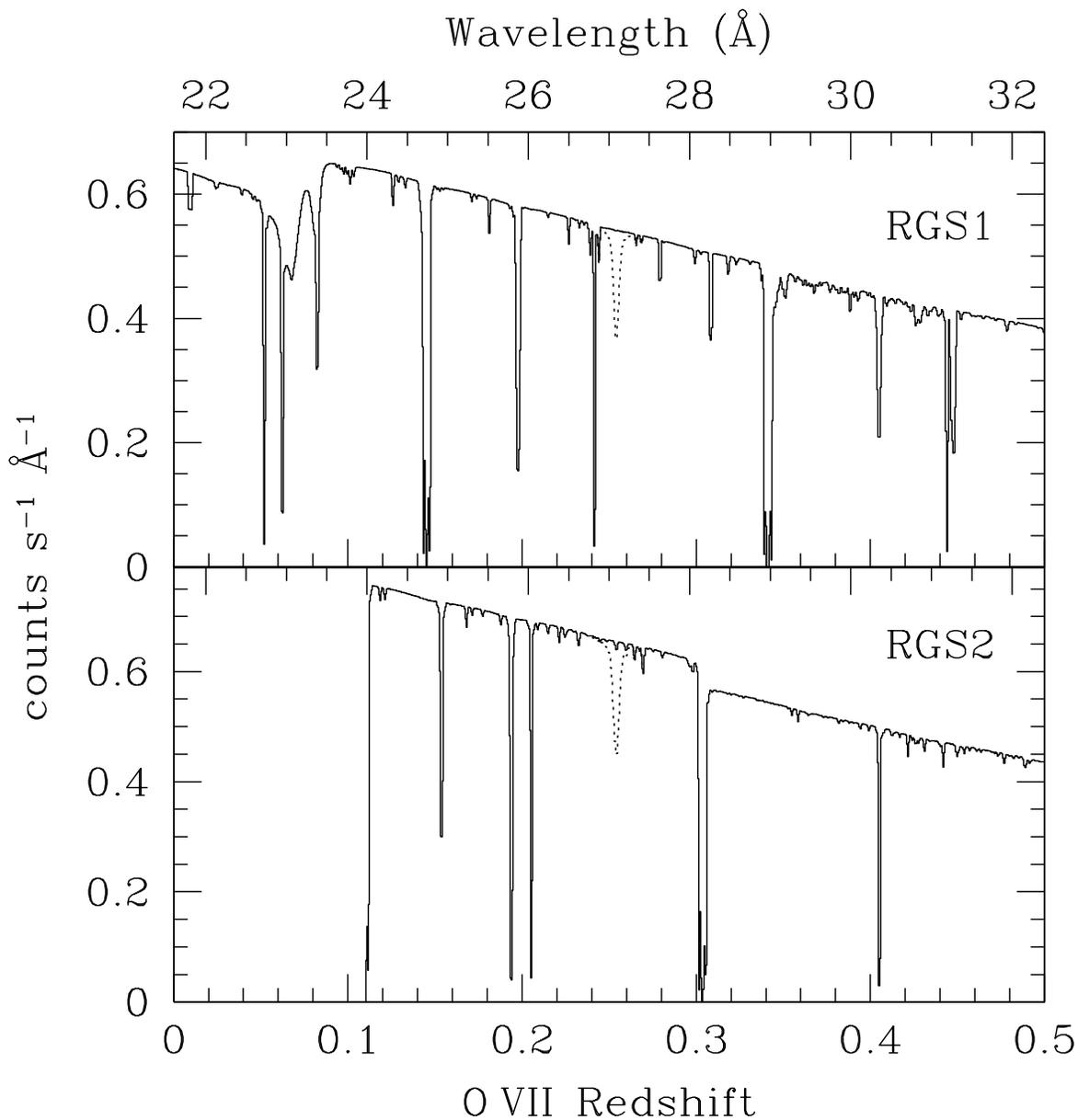}
\caption{RGS1 (top panel) and RGS2 (bottom panel) instrumental response
models for the \ovii\ $z=0-0.5$ region, as a function of wavelength
(upper axes) and redshift relative to $\lambda=21.602$\AA\ (lower axes);
a strong $z=0.25$ line with $N_{\rm OVII}=10^{16}$\,cm$^{-2}$ 
($W_\lambda=36$\,m\AA) is shown for reference.
\label{fig_rgsmod}}
\end{figure}

\begin{figure}
\plotone{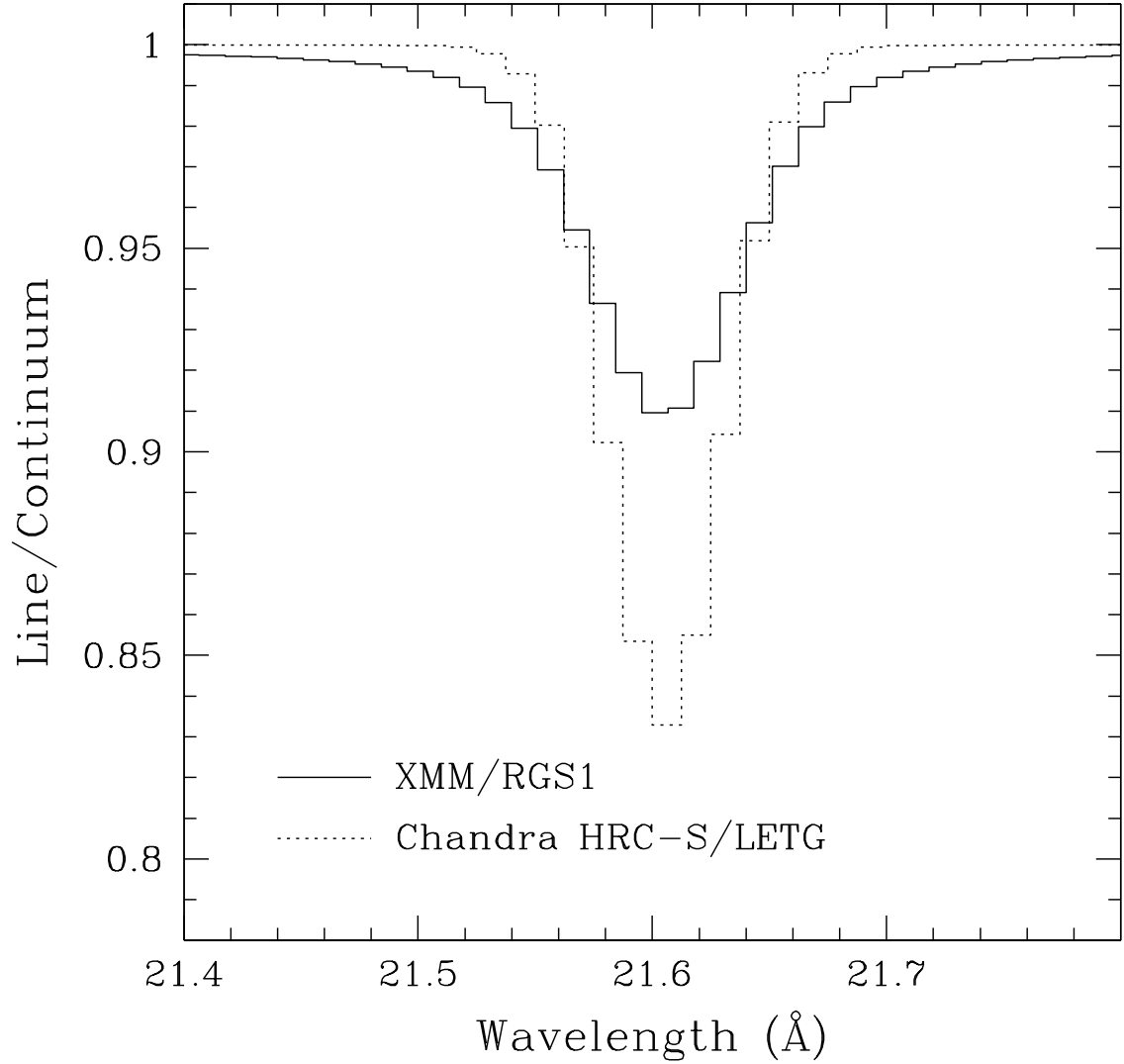}
\caption{Comparison of the \xmm\ RGS1 (solid) and \chandra\ HRC-S/LETG
(dotted) line spread functions for a $W_\lambda=10$\,m\AA\ unresolved 
absorption line at 21.602\,\AA. 
\label{fig_lsfplot}}
\end{figure}

Why, then, with $2-4$ times the counts per resolution element, was
\xmm\ unable to detect the intervening absorption systems seen by
\chandra?  Several factors appear to have been involved in this
non--detection, primarily: (1) narrow instrumental features caused by
bad detector columns, (2) the broader LSF as compared
to \chandra/LETG, and (3) fixed--pattern noise at long wavelengths: 
\begin{enumerate}
\item{While broad instrumental features can be taken into
account by modifications to the continuum model (as in N05), it is 
difficult or impossible
to distinguish narrow features from true absorption lines; thus, any line
falling near one of the detector features shown in Figure~\ref{fig_rgsmod} 
can easily be lost\footnote{These response file data can be found at 
\url{http://www.astronomy.ohio-state.edu/$\sim$smita/xmmrsp/}}.
This was responsible for the non-detection of the $z=0.011$ \ovii\ K$\alpha$
line.  Although it was the strongest line reported by N05,
its wavelength falls directly on a narrow RGS1 feature and within
the non--functional CCD4 region on RGS2,
thereby preventing this line from being detectable with either RGS.  Since
18\% of the wavelength space for studying redshifted \ovii\ 
($\lambda>21.6$\,\AA) toward Mrk 421 is directly blocked by these narrow 
features
(with this number climbing to about 60\% if resolution elements immediately
adjacent to bad columns are included), these bad columns present the
single greatest hindrance to \xmm/RGS studies of the WHIM.}
\item{Even for lines where both RGS1 and RGS2 data are available and the 
instrumental response appears to be relatively smooth, the lower resolution
of \xmm\ contributes to the nondetectability of the
weaker $z>0$ absorption lines.  Figure~\ref{fig_lsfplot} shows the
LSFs for both \xmm/RGS1 (solid) and \chandra/LETG assuming an unresolved
line with $W_\lambda=10$\,m\AA\ at 21.6\,\AA.  While the core of the RGS1 
response is 
$\sim 20$\% broader than that of the LETG, the RGS1 LSF
has extremely broad wings: only 68\% of the line flux is contained within
the central 0.1\AA\ of the RGS1 LSF, as opposed to 96\% for the LETG. 
This reduces the apparent depth of absorption lines by about a factor of
two as compared to \chandra/LETG, severely decreasing the line detectability.}
\item{At long wavelengths ($\lambda\ga 29$\,\AA)
strong fixed--pattern noise is apparent as a sawtooth pattern in the
instrumental response, strongly impeding the detection of species
such as \nvi\ and \cvi.  Indeed, in these wavelength regimes
(the lower two panels of Figure~\ref{fig_xmmspec}), the \nvi\ and \cvi\
absorption lines are nearly indistinguishable from the continuum.}
\end{enumerate}

\section{Conclusion}
We have presented the highest signal--to--noise coadded \xmm\ grating 
spectrum of 
Mrk 421 to date, incorporating all available archival data.  This 
spectrum serves as an independent check on the recent detection of 
two $z>0$ WHIM filaments by N05.  While none of the \chandra--detected
absorption lines are seen in the \xmm\ spectrum, the upper limits derived
from the \xmm\ data are consistent with the equivalent widths
reported by N05 (even though the \xmm\ data contain a larger number
of counts), and hence do not jeopardize the validity of the \chandra\
measurement.  The non--detections can be attributed primarily to narrow
instrumental features in RGS1 and RGS2, as well as the inferior spectral
resolution of \xmm\ and fixed--pattern noise at longer wavelengths.
This underscores the extreme difficulty 
of detecting the WHIM, illustrates how the aforementioned (apparently small) 
effects can greatly affect the delicate measurement 
of weak absorption lines, and re--emphasizes the importance of high resolution
and a smooth instrumental response function for current and future WHIM
absorption line studies.

\acknowledgments
We thank the \xmm\ team for their efforts on this excellent mission,
the helpdesk staff for their assistance with the data reduction, and
the anonymous referee for helpful comments.
This research is based on archival data obtained with XMM-Newton, an ESA 
science mission with instruments and contributions directly funded by ESA 
Member States and NASA.  RJW is supported by an Ohio State University 
Presidential Fellowship, and FN acknowledges the support of NASA Long--Term
Space Astrophysics Grant NNG04GD49G.




\clearpage
\begin{deluxetable}{lcccc}
\tabletypesize{\footnotesize}
\tablecolumns{5}
\tablewidth{240pt}
\tablecaption{\xmmnewton\ observation log \label{tab_log}}
\tablehead{
\colhead{ID} &
\colhead{Date} &
\colhead{$t_{\rm exp}$\tablenotemark{a}} &
\colhead{$t_{\rm filt}$\tablenotemark{b}} &
\colhead{Rate\tablenotemark{c}} 
\\
\colhead{} &
\colhead{} &
\colhead{ks} &
\colhead{ks} &
\colhead{s$^{-1}$} 
}

\startdata
0099280101 &2000 May 25 &63.8 &21.2 &15.7\\
0099280201 &2000 Nov 01 &40.1 &34.1 &5.4\\
0099280301 &2000 Nov 13 &49.8 &46.6 &15.3\\
0099280501 &2000 Nov 13 &21.2 &17.8 &17.2\\
0136540101 &2001 May 08 &38.8 &36.1 &11.7\\
0136540301 &2002 Nov 04 &23.9 &20.5 &11.7\\
0136540401 &2002 Nov 04 &23.9 &20.1 &13.6\\
0136540701 &2002 Nov 14 &71.5 &62.8 &16.4\\
0136541001 &2002 Dec 01 &70.0 &58.1 &8.3\\
0158970101 &2003 Jun 01 &43.0 &25.3 &9.0\\
0158970201 &2003 Jun 02 &9.0  &6.6  &9.7\\
0158970701 &2003 Jun 07 &48.9 &29.9 &5.4\\
0158971201 &2004 May 06 &65.7 &40.5 &19.5\\
0162960101 &2003 Dec 10 &30.0 &17.5 &9.8\\
\hline\hline
TOTAL      &             &572.3 &437.1 &12.2\\
\enddata
\tablenotetext{a}{Total observation duration.}
\tablenotetext{b}{Effective RGS1 exposure time after filtering for periods
of high background levels.}
\tablenotetext{c}{Average count rate in the filtered RGS1 first--order source
spectral extraction region.}
\end{deluxetable}

\clearpage
\begin{deluxetable}{lccccccc}
\tabletypesize{\footnotesize}
\tablecolumns{8}
\tablecaption{Absorption line equivalent width measurements \label{tab_ew}}
\tablehead{
\colhead{Line} &
\colhead{$\lambda$\tablenotemark{a}} &
\colhead{$z$\tablenotemark{a}} &
\colhead{$W_{\lambda, {\rm N05a}}$\tablenotemark{a}} &
\colhead{$W_{\lambda, {\rm R1}}$\tablenotemark{b}} &
\colhead{$W_{\lambda, {\rm R2}}$\tablenotemark{b}} &
\colhead{$W_{\lambda, {\rm R1+R2}}$\tablenotemark{b}} &
\colhead{Note}
\\
\colhead{} &
\colhead{\AA} &
\colhead{} &
\colhead{m\AA} &
\colhead{m\AA} &
\colhead{m\AA} &
\colhead{m\AA} &
\colhead{}
}

\startdata
\ion{Ne}{9}$_{K\alpha}$ &$13.80\pm 0.02$ &$0.026\pm 0.001$ &$<1.5$ &$<5.2$ &$<1.9$ &$<2.9$ &1\\
\ion{O}{7}$_{K\beta}$ &$19.11\pm 0.02$ &$0.026\pm 0.001$ &$<1.8$ &$<2.5$ &$<2.1$ &$<1.5$ &\\
\ion{O}{8}$_{K\alpha}$ &$19.18\pm 0.02$ &$0.011\pm 0.001$ &$<4.1$ &$<7.6$ &$<5.8$ &$<4.1$ &\\
\ion{O}{8}$_{K\alpha}$ &$19.48\pm 0.02$ &$0.027\pm 0.001$ &$<1.8$ &\nodata &$<3.9$ &\nodata &2\\
\ion{O}{7}$_{K\alpha}$ &$21.85\pm 0.02$ &$0.011\pm 0.001$ &$3.0^{+0.9}_{-0.8}$ &\nodata &\nodata &\nodata &2,3\\
\ion{O}{7}$_{K\alpha}$ &$22.20\pm 0.02$ &$0.028\pm 0.011$ &$2.2\pm 0.8$ &$<3.9$ &\nodata &\nodata &3\\
\ion{N}{7}$_{K\alpha}$ &$25.04\pm 0.02$ &$0.010\pm 0.001$ &$1.8\pm 0.9$ &$<3.0$ &$<6.0$ &$<4.4$ &\\
\ion{N}{7}$_{K\alpha}$ &$25.44\pm 0.02$ &$0.027\pm 0.001$ &$3.4\pm 1.1$ &$<4.3$ &$<4.2$ &$<3.5$\\
\ion{N}{6}$_{K\alpha}$ &$29.54\pm 0.02$ &$0.026\pm 0.001$ &$3.6\pm 1.2$ &$<3.8$ &$<8.7$ &$<3.4$ &\\
\ion{C}{6}$_{K\alpha}$ &$34.69\pm 0.02$ &$0.028\pm 0.001$ &$2.4\pm 1.3$ &$<5.5$ &$<5.2$ &$<4.2$ &\\
\enddata
\tablenotetext{a}{\ Line wavelength, redshift, and equivalent width measurements 
(or $3\sigma$ upper limits) from \citet{nicastro05a}.}
\tablenotetext{b}{$2\sigma$ equivalent width upper limits measured from the
RGS1 only (R1), RGS2 only (R2), and joint (R1+R2) fits to the XMM--Newton
spectrum, when available.}
\tablecomments{
(1) A nearby chip gap in RGS1 renders this measurement unreliable, so only 
the RGS2 measurement was used in Figure~\ref{fig_xmmspec};
(2) Line was unmeasurable in RGS1 because of a detector feature;
(3) Line was unmeasurable in RGS2 because of a detector feature.
}
\end{deluxetable}

\end{document}